\RequirePackage{fix-cm}
\documentclass[twocolumn]{svjour3}          
\smartqed  
\usepackage{graphicx}
\usepackage{subfigure}
\usepackage{amssymb}
\usepackage{arydshln}
\usepackage{amsmath}
\usepackage[boxed]{algorithm2e}
\usepackage{mathrsfs}
\usepackage{multicol}
\usepackage{cite}
\journalname{Comp. Mech.}

\begin{document}

\title{A consistent interface element formulation for geometrical and material nonlinearities
\thanks{} }


\author{J. Reinoso         \and
        M. Paggi
}


\institute{J. Reinoso \at
              IMT Institute for Advanced Studies Lucca, Piazza San Francesco 19, 55100 Lucca, Italy\\
              Institute of Structural Analysis. Leibniz Universit\"at  Hannover, Appelstr. 9A, 30167 Hannover, Germany\\
              Group of Elasticity and Strength of Materials, School of Engineering, University of Seville, Camino de los Descubrimientos s/n, 41092, Seville, Spain\\
              \email{j.reinoso@isd.uni-hannover.de}           
           \and
           M. Paggi \at
              IMT Institute for Advanced Studies Lucca, Piazza San Francesco 19, 55100 Lucca, Italy\\
              \email{marco.paggi@imtlucca.it}
}

\date{Received: date / Accepted: date}

\maketitle

\begin{abstract}
Decohesion undergoing large displacements takes place in a wide
range of applications. In these problems, interface element
formulations for large displacements should be used to accurately
deal with coupled material and geometrical nonlinearities. The
present work proposes a consistent derivation of a new interface
element for large deformation analyses. The resulting compact
derivation leads to a operational formulation that enables the
accommodation of any order of kinematic interpolation and
constitutive behavior of the interface. The derived interface
element has been implemented into the finite element codes FEAP and
ABAQUS by means of user-defined routines. The interplay between
geometrical and material nonlinearities is investigated by
considering two different constitutive models for the interface
(tension cut-off and polynomial cohesive zone models) and small or
finite deformation for the continuum. Numerical examples are
proposed to assess the mesh independency of the new interface
element and to demonstrate the robustness of the formulation. A
comparison with experimental results for peeling confirms the
predictive capabilities of the formulation.

\vspace{1em}
Notice: this is the author's version of a work that was
accepted for publication in Computational Mechanics. Changes
resulting from the publishing process, such as editing, structural
formatting, and other quality control mechanisms may not be
reflected in this document. A definitive version was published in
Computational Mechanics, Vol. 54 (2014) 1569–1581,
DOI:10.1007/s00466-014-1077-2

\end{abstract}

\section{Introduction}
In recent years, cohesive zone models (CZMs) have been used in a
variety of engineering applications concerning the formation of free
surfaces due to the development of fracture processes. Relying on
the seminal work of Barenblatt \cite{giB}, CZMs have been massively
incorporated into computational frameworks, especially in the
context of the nonlinear finite element (FE) method, as a
consequence of two primary reasons: (i) the high versatility of the
approach to accommodate different phenomenological fracture events,
and (ii) the relative simplicity to numerically implement interface
elements as user defined subroutines into research and commercial FE
codes. In this context, the basic ingredient that characterizes CZMs
is the so-called nonlinear traction-displacement jump relationship
which relates the cohesive tractions to the relative opening and
sliding displacements at the interface, where various contributions
have been proposed, see \cite{EGGP,ngu,zamm,PWCM} for a wide review
of formulations and a recent special issue on the topic.

Applications cover several fields and range from quasi-static
fracture in quasi-brittle solids \cite{aH,AC} with special attention
to modelling snap-back instabilities during crack propagation
\cite{AC1,AC2}, crack propagation in composites
\cite{allix,camanho,reinoso}, coupled thermo-mechanical applications
\cite{HS,OBG,SP}, micromechanical and multi-scale analyses
\cite{BSG08,PW11,Petal}, fracture and contact at interfaces
\cite{CPZ}, combination of friction and cohesive fracture at
interfaces \cite{sacco1,borino}, and flaw-tolerance assessment in
bio-inspired materials \cite{gao,PW12}, among others.

Specific contributions related to finite elements regarded the study
of the effect of the numerical integration of interface elements
\cite{SB}, ill-conditioning situations \cite{pande}, convergence
issues \cite{alfano,borst} and the use of a set of overlapping
cohesive segments \cite{remmers}.

In applications regarding thin structural elements subjected to
large displacements, as, e.g., in biological membranes, paper
sheets, elastomers, viscoelastic materials for the encapsulation of
solar cells, the complexity relies on the fact that during the
simulation the deformed configuration cannot be approximated by the
underformed one due to the occurrence of large displacements.
Therefore, the computation of the interface gap (global or projected
over a local reference basis) according to the initial underformed
geometry can lead to errors depending upon the specific applications
and materials tested. Thus, large-displacement analyses require
tracking of the surface separation, the relative rotations between
the two sides of the interface and the simultaneous deformation of
the two bodies separated by the interface. A pioneering attempt to
solve this problem is due to Ortiz and Pandolfi \cite{OP99}, who
suggested the adoption of a reference middle surface of the cohesive
element in the current configuration to define a convenient
(deformed) surface for the calculation of the normal and tangential
directions to the interface. Nevertheless, their resulting
formulation specified for a quadratic 3D interface element for
matching tetrahedra and stemming from the differentiation of the
cohesive tractions with respect to the normal unit vector to the
middle surface led to a non-symmetric geometric stiffness matrix. In
\cite{RRD02}, a 3D large displacement interface element was used
based on the aforementioned formulation in order to simulate
standard fracture mechanics tests in thin aluminum panels. In that
case, a residual with a rotation matrix updated along the
deformation process was considered, whereas its consistent
linearization did not take into account the dependence of the
standard $\mathbf{B}$-operator with respect to the kinematic field.

In \cite{QCA01}, an alternative formulation for a 2D interface
element in large displacements was proposed by introducing a non
symmetric co-rotational reference system coincident with one of the
two deformed sides of the interface. As also admitted by the
authors, this co-rotational description leads to a very complex
formulation of cumbersome implementation. Approximate 2D and 3D
formulations with emphasis on the problem of interface fibrilation
were recently proposed in \cite{BSG07,BSG08b}. In this instance, the
kinematics of the interface element was assumed to be like that of
2D or 3D trusses under large displacements and rotations. More
recently, an interface element in large displacements for fully
coupled thermo-mechanical applications was proposed in \cite{FBK13}.
The authors defined the CZM relation in a global reference system,
similarly to the method proposed in \cite{BSG07}, but did not
consider a CZM relation that takes into account the contribution of
different fracture modes. Although this could be an advantage to
simplify the burn of the linearization of the residual, actually it
requires the use of integrated formulations to deal with the
transition from small to large displacement regimes as suggested in
\cite{BSG07}. Moreover, to the present authors' view, the
geometrical contribution to the stiffness matrix was not clearly
addressed in \cite{FBK13}.

The objective of this paper is concerned with the development of a
consistent interface element formulation for material and
geometrical nonlinearities and the derivation of its corresponding
finite element implementation. The starting point of the consistent
derivation is the analysis of the interface contribution to the
Principle of Virtual Work of the whole mechanical system, its
virtual variation, discretization and then linearization. As shown
in the next sections, the resulting derivation leads to a simple and
compact operational formulation in which the geometric and the
material contribution to the element stiffness matrix are clearly
identified. In addition to this, one of the most appealing aspects
of the model herein proposed relies on its  versatility to
accommodate any 2D and 3D finite element typologies along with any
interface decohesion law, without any lack of generality.

The article is organized as follows. In Section 2, the governing
equations of the large displacement interface formulation and the
corresponding finite element discretization are established. The
constitutive models for the bulk and for the interface used in this
investigation are then briefly outlined in Section 3. In particular,
a tension cut-off model and a polynomial CZM are considered as two
limit cases representative for very brittle or ductile interface
performances. Section 4 addresses the main issues regarding the FE
implementation in the context of the classical iterative
Newton-Rapshon solution scheme. Section 5 presents a series of test
problems, applications to peeling and proves the robustness of the
formulation and its ability to capture experimental results related
to peeling tests of very thin layers. Finally, the main conclusions
are given in Section 6.

\section{Large displacement interface model and finite element formulation}

\subsection{Variational framework}

\begin{figure}
\centering
\includegraphics[width=.45\textwidth,angle=0]{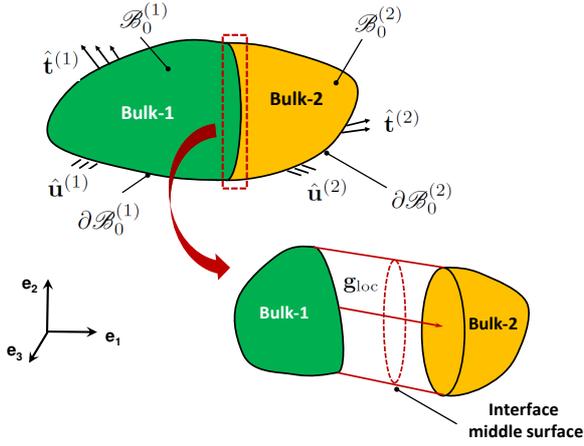}
\caption{A schematic definition of two bodies separated by a cohesive interface.}\label{fig1}
\end{figure}

The point of departure of the present formulation relies on the
interface contribution to the expression of the Principle of Virtual
Work of the whole system. Let us to assume two deformable bodies
$\mathscr{B}_{0}^{(i)}$ $(i=1,2)$ in the reference configuration
(denoted as Bulk-1 and Bulk-2 in Fig.\ref{fig1}), which could have
different constitutive relations that characterize their mechanical
performance. As customary, both bodies are subjected to the external
body forces $\mathbf{F}_{v}^{(i)}$ $(i=1,2)$. The boundary
conditions applied on their boundaries are
$\mathbf{t}^{i}=\hat{\mathbf{t}}^{i}$ on $\partial
\mathscr{B}_{0,\mathbf{t}}^{i}$ and
$\mathbf{u}^{i}=\hat{\mathbf{u}}^{i}$ on $\partial
\mathscr{B}_{0,\mathbf{u}}^{i}$ $(i=1,2)$.

The bodies undergo a motion $\boldsymbol \phi:\mathscr{B}_{0} \times
[0,t] \rightarrow \mathscr{R}^{3}$, where $[0,t]$ is the time step
interval, that maps the reference material points ($\mathbf{X} \in
\mathscr{B}_{0}$) onto the current material points ($\mathbf{x} \in
\mathscr{B}$), such that $\mathbf{x}=\boldsymbol \phi(\mathbf{X},
t)$. The deformation gradient of the transformation is defined as
$\mathbf{F}:=\partial_{\mathbf{X}} \phi(\mathbf{X}, t)$, with the
Jacobian $J = \text{det} [\mathbf{F}]$ and $\partial_{\mathbf{X}}$
denoting the partial derivative with respect to the reference frame.
Moreover, it is supposed that the interface between both solids is
characterized by the presence of a cohesive surface $S_{0}$.

Focusing our attention on the analysis of the interface between the
solids, the contribution of the interface cohesive tractions
$\mathbf{T}$, acting on $S_{0}$, to the Principle of Virtual Work of
the mechanical system in the reference configuration is:
\begin{equation}\label{eq1}
\Pi_{\mathrm{intf}} (\mathbf{g}_{\mathrm{loc}}) =\int_{S_{0}} \mathbf{g}_{\mathrm{loc}}^{\mathrm{T}}
\mathbf{T}\,\mathrm{d}S
\end{equation}
where $\mathbf{g}_{\mathrm{loc}}$ is the gap vector accounting for
opening and sliding displacements between the two sides of the
interface. Note that, due to the geometrical nonlinearity, the
traction vector previously defined corresponds to the nominal first
Piola-Kirchhoff tractions related to the local basis of the
interface in the reference configuration.

It is worth noting that in the large deformation setting, the gaps
vector vanishes when the body undergoes rigid body motions, thus
confirming the frame indifference of the formulation proposed in
this paper.

The virtual variation of $\Pi_{\mathrm{intf}}$ according to the
principle of virtual displacements reads:
\begin{equation}
\begin{aligned}\label{eq2}
\delta\Pi_{\mathrm{intf}} (\mathbf{g}_{\mathrm{loc}}) =&\int_{S_{0}}
\left(\dfrac{\partial\mathbf{g}_{\mathrm{loc}}}{\partial\mathbf{u}}
\delta\mathbf{u}\right)^{\mathrm{T}}\,\mathbf{T}\mathrm{d}S\\
=&\delta\mathbf{u}^{\mathrm{T}}\int_{S_{0}}
\left(\dfrac{\partial\mathbf{g}_{\mathrm{loc}}}{\partial\mathbf{u}}\right)^{\mathrm{T}}\,\mathbf{T}\mathrm{d}S
\end{aligned}
\end{equation}

\begin{figure}
\centering
\includegraphics[width=.45\textwidth,angle=0]{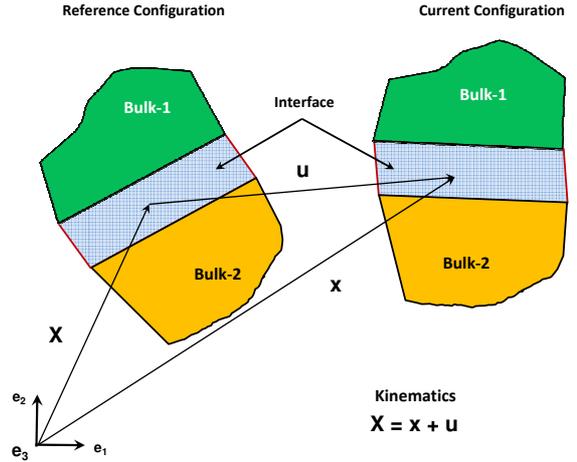}
\caption{Kinematics definition of the interface along the deformation process.}\label{fig2}
\end{figure}

In case of large displacements, the updated coordinates
of a generic point are given by $\mathbf{x}=\mathbf{X}+\mathbf{u}$, see Fig.\ref{fig2}.

\begin{figure}
\centering
\includegraphics[width=.45\textwidth,angle=0]{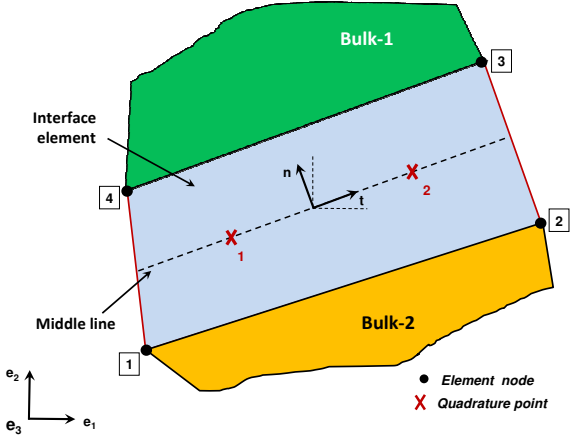}
\caption{Sketch of the interface element with node numbering and integration points.}\label{fig3}
\end{figure}

As is generally proposed for interface formulations, it is
convenient to define a middle line (in the 2D case) in the updated
configuration by averaging the position vectors and the displacement
fields of the upper and lower sides of the interface, see
Fig.\ref{fig3} after performing a standard discretization process.
Hence, the position vector $\overline{\mathbf{x}}$ of a generic
point along this middle line can be determined by pre-multiplying
the positioning vector $\mathbf{x}$ by an averaging operator
$\mathbf{M}$:
\begin{equation}\label{eq3}
\overline{\mathbf{x}}=\mathbf{M}\mathbf{x}
\end{equation}

\subsection{Finite element formulation}

Based on isoparametric interpolation, the position vector at the
interface can be approximated through:
\begin{equation}\label{eq4a}
\mathbf{x} \cong \overline{\mathbf{x}}^e =  \mathbf{N}\mathbf{x}^n
\end{equation}
where $\mathbf{x}^n$ denotes the nodal position vector (the
superscript $n$ identifies nodal quantities), and $\mathbf{N}$ is
the the operator that collects the shape functions and it depends on
the natural coordinate of the element $\xi$. Introducing now the
discretization of the interface into Eq.\eqref{eq3}, the
interpolated average position vector yields:
\begin{equation}\label{eq4}
\overline{\mathbf{x}}\cong
\overline{\mathbf{x}}^e=\mathbf{N}\mathbf{M}\mathbf{x}^n
\end{equation}

Similarly, the coordinates of the points belonging to the middle line in the reference configuration,
$\overline{\mathbf{X}}$, and their displacement vector, $\overline{\mathbf{u}}$, can be computed via a
standard interpolation procedure from the nodal quantities
\begin{equation}\label{eq5}
\overline{\mathbf{X}}\cong\overline{\mathbf{X}}^e=\mathbf{N}\mathbf{M}\mathbf{X}^n,\quad
\overline{\mathbf{u}}\cong\overline{\mathbf{u}}^e=\mathbf{N}\mathbf{M}\mathbf{d}
\end{equation}
where $\mathbf{X}^n$ and $\mathbf{d}$ denote the position vector of
the nodes in the reference configuration and their nodal
displacement vector, respectively.

In 2D, the tangential and the normal vectors $\mathbf{t}$ and
$\mathbf{n}$ to the middle line of the interface element used to
define the local frame are given by:
\begin{equation}\label{eq6b}
\mathbf{t}=\dfrac{\partial
\overline{\mathbf{x}}^e}{\partial \xi},\quad
\mathbf{n}\cdot\mathbf{t}=0
\end{equation}
Note that in 3D application, in line with the derivation proposed in
Ortiz and Pandolfi \cite{OP99}, the convective tangential and normal
vectors to the middle surface of the interface element
($\mathbf{t}_1$, $\mathbf{t}_2$ and $\mathbf{n}$) used to define the
local frame are determined via differentiation of the average
coordinates with respect to the natural coordinates $\xi$ and
$\eta$:
\begin{equation}\label{eq6a}
\mathbf{t}_1=\dfrac{\partial \overline{\mathbf{x}}^e}{\partial
\xi},\quad \mathbf{t}_2=\dfrac{\overline{\mathbf{x}}^e}{\partial
\eta},\quad \mathbf{n}=\mathbf{t}_1\times\mathbf{t}_2
\end{equation}

The gap vector in the reference cartesian frame, $\mathbf{g}$, can
be obtained by pre-multiplying the nodal displacement vector
$\mathbf{d}$ by a suitable operator $\mathbf{L}$ which provides the
difference between the displacements of the upper and the lower
bodies at the interface. Accordingly, within the FE discretization
we have:
\begin{equation}\label{eq7}
\mathbf{g}\cong\mathbf{g}^e=\mathbf{N}\mathbf{L}\mathbf{d}
\end{equation}
The constitutive relation for the interface, i.e., the so-called cohesive zone model (CZM), is usually provided in
a local frame defined by the normal and the tangential
vectors to the average line of the interface element in order to distinguish between fracture Modes I and II, as introduced
in Eq.\eqref{eq6b}. Therefore, the gap vector in this local frame,
$\mathbf{g}_{\mathrm{loc}}$, has to be computed by multiplying the gap vector in the reference frame by a rotation operator
$\mathbf{R}$:
\begin{equation}\label{eq8}
\mathbf{g}_{\mathrm{loc}}=\mathbf{R}(\mathbf{u})\mathbf{g}
\end{equation}
It is remarkable to note that, in case of large displacements, the
operator $\mathbf{R}(\mathbf{u})$ is a function of the displacement
field. Its expression is detailed in Section 4 for the 2D case that
represents the main scope of the present work (the 3D version can be
derived adapting the formulation here developed). Consequently, a
consistent formulation must take into account this dependency in the
subsequent linearization of the discretized version of the interface
contribution to the Principle of Virtual Work within the classical
Newton-Raphson iterative solution scheme. This dependency will lead
to the so-called geometric contribution to the element stiffness
matrix. Introducing the FE discretization, Eq.\eqref{eq8} can be
rewritten as:
\begin{equation}\label{eq9}
\mathbf{g}_{\mathrm{loc}}^{e}=\mathbf{R}(\mathbf{d})\mathbf{N}\mathbf{L}\mathbf{d}
\end{equation}

Examining the terms entering the virtual variation of the virtual
work in Eq.\eqref{eq2}, the partial derivative
$(\partial\mathbf{g}_{\mathrm{loc}}/\partial\mathbf{u})$ is
approximated by:
\begin{equation}\label{eq11}
\dfrac{\partial\mathbf{g}_{\mathrm{loc}}}{\partial\mathbf{u}}
\cong
\dfrac{\partial\mathbf{g}_{\mathrm{loc}}^e}{\partial\mathbf{d}}=
\mathbf{R}(\mathbf{d})\mathbf{N}\mathbf{L}+\dfrac{\partial
\mathbf{R}(\mathbf{d})}{\partial\mathbf{d}}\mathbf{N}\mathbf{L}\mathbf{d}
\end{equation}
where the differentiation of the second order tensor $\mathbf{R}$
with respect to the components of the vector $\mathbf{d}$ leads to a
third order tensor. Note that in \eqref{eq11} the formulation is
simplified by omitting the second derivative of the rotation matrix
with respect to the displacement vector. This vanishes in case of
linear displacement interpolation under the assumption that the norm
of the tangent vector $\mathbf{t}$ does not depend on the
displacement field. In this regard, we also assessed the role of
this term based on representative numerical tests adopting
alternative interpolation schemes. It was found that this term has
an almost negligible effect on the results and it is therefore
reasonable to be neglected.

The operator $\mathbf{B}=\mathbf{N}\mathbf{L}$ is now introduced and
Eq.\eqref{eq11} can be rephrased as:
\begin{equation}\label{eq12}
\dfrac{\partial\mathbf{g}_{\mathrm{loc}}^e}{\partial\mathbf{d}}=
\mathbf{R}\mathbf{B}+\dfrac{\partial \mathbf{R}}{\partial
\mathbf{d}}\mathbf{B}\mathbf{d}
\end{equation}
The matrices $\mathbf{R}$ and $\mathbf{B}$ are evaluated at the
element level, though the typical superscript ($e$) has been omitted
here to simplify notation.

Inserting this intermediate result into the discretized version of
Eq.\eqref{eq2}, where $\mathbf{u}$ is simply replaced by
$\mathbf{d}$, the following general formulation valid for any kind
of interface element topology dealing with geometric and material
nonlinearities is derived:
\begin{equation}\label{eq14}
\begin{aligned}
&\delta\Pi_{\mathrm{intf}}^e =
\delta\mathbf{d}^{\mathrm{T}}\int_{S_0}
\left(\mathbf{R}\mathbf{B}+\dfrac{\partial \mathbf{R}}{\partial
\mathbf{d}}\mathbf{B}\mathbf{d}\right)^{\mathrm{T}}\mathbf{T}\,\mathrm{d}S
\end{aligned}
\end{equation}

The solution of the variational equation
$\delta\Pi_{\mathrm{intf}}^e=\delta\mathbf{d}^{\mathrm{T}}\mathbf{f}_{\mathrm{intf}}^e=0$
$\forall \delta\mathbf{d}$ results in the equations set
$\mathbf{f}_{\mathrm{intf}}^e=0$, where
$\mathbf{f}_{\mathrm{intf}}^e$ is a nonlinear function of the
unknown $\mathbf{d}$ and assumes the role of the residual vector in
the Newton-Raphson iterative scheme:
\begin{equation}\label{residual}
\mathbf{f}^{e,k}_{\mathrm{intf}}=\int_{S_0}
\left(\mathbf{R}\mathbf{B}+\dfrac{\partial \mathbf{R}}{\partial
\mathbf{d}}\mathbf{B}\mathbf{d}\right)^{\mathrm{T}}\mathbf{T}\,\mathrm{d}S
\end{equation}
which leads to the following equations set for the computation of
the corrector $\Delta \mathbf{d}$ at each iteration $k$:
\begin{subequations}
\begin{align}
\mathbf{K}^{e,k}\Delta \mathbf{d}&=-\mathbf{f}_{\mathrm{intf}}^{e,k}\\
\mathbf{d}^{k+1}&=\mathbf{d}^{k}+\Delta \mathbf{d}
\end{align}
\end{subequations}
To alleviate the notation, the superscript $k$ will be omitted in
the sequel. Following standard arguments of nonlinear FE
formulations, the element stiffness matrix $\mathbf{K}^{e}$ is
obtained from the linearization of the residual, i.e.,
$\mathbf{K}^{e}=\partial \mathbf{f}_{\mathrm{intf}}/\partial
\mathbf{d}$ and it is evaluated by using the displacement field
solution at the iteration $k$:
\begin{equation}\label{eq16}
\begin{aligned}
\mathbf{K}^{e}=&\int_{S_0}
\left[2\mathbf{B}^{\mathrm{T}}\dfrac{\partial\mathbf{R}^{\mathrm{T}}}{\partial
\mathbf{d}}\mathbf{T}\right.\\
&\left.+\left(\mathbf{B}^{\mathrm{T}}\mathbf{R}^{\mathrm{T}}+\mathbf{d}^{\mathrm{T}}\mathbf{B}^{\mathrm{T}}\dfrac{\partial\mathbf{R}^{\mathrm{T}}}{\partial
\mathbf{d}}\right)\dfrac{\partial \mathbf{T}}{\partial
\mathbf{d}}\right]\,\mathrm{d}S
\end{aligned}
\end{equation}
In this derivation, as it was already stated, the second-order
differentiation of the rotation matrix $\mathbf{R}$ was omitted.

The derivative of the cohesive traction vector $\mathbf{T}$ with
respect to the displacement vector $\mathbf{d}$ can be determined
via a chain rule differentiation:
\begin{equation}\label{eq17}
\begin{aligned}
\dfrac{\partial \mathbf{T}}{\partial \mathbf{d}}=\dfrac{\partial
\mathbf{T}}{\partial \mathbf{g}_{\mathrm{loc}}}
\dfrac{\mathbf{g}_{\mathrm{loc}}}{\partial
\mathbf{d}}=\mathbf{C}\left(\mathbf{R}\mathbf{B}+\dfrac{\partial
\mathbf{R}}{\partial \mathbf{d}}\mathbf{B}\mathbf{d}\right)
\end{aligned}
\end{equation}
where
$\mathbf{C}=\dfrac{\partial\mathbf{T}}{\partial\mathbf{g}_{\mathrm{loc}}}$
represents the tangent interface constitutive matrix whose
expression will be detailed in the next section.

After some algebra we obtain the following result:
\begin{equation}\label{eq18}
\begin{aligned}
\mathbf{K}^{e}=&
\int_{S_0} \mathbf{B}^{\mathrm{T}}\mathbf{R}^{\mathrm{T}}\mathbf{C}\mathbf{R}\mathbf{B}\,\mathrm{d}S\\
&+\int_{S_0}
\left[
2\mathbf{B}^{\mathrm{T}}\dfrac{\partial\mathbf{R}^{\mathrm{T}}}{\partial
\mathbf{d}}\mathbf{T}+\mathbf{d}^{\mathrm{T}}\mathbf{B}^{\mathrm{T}}\dfrac{\partial\mathbf{R}^{\mathrm{T}}}{\partial\mathbf{d}}\mathbf{C}\dfrac{\partial\mathbf{R}}{\partial\mathbf{d}}\mathbf{B}\mathbf{d}\right.\\
&\left.+\mathbf{B}^{\mathrm{T}}\mathbf{R}^{\mathrm{T}}\mathbf{C}\dfrac{\partial\mathbf{R}}{\partial\mathbf{d}}\mathbf{B}\mathbf{d}+\mathbf{d}^{\mathrm{T}}\mathbf{B}^{\mathrm{T}}\dfrac{\partial\mathbf{R}^{\mathrm{T}}}{\partial\mathbf{d}}\mathbf{C}\mathbf{R}\mathbf{B}\right]\,\mathrm{d}S
\end{aligned}
\end{equation}

Summarizing, the tangent stiffness matrix which accounts for both
the material and the geometric contributions reads:
\begin{subequations}\label{eq19}
\begin{align}
&\mathbf{K}^e=\mathbf{K}^e_{\mathrm{mat}}+\mathbf{K}^e_{\mathrm{geom}}\label{19a}\\
&\mathbf{K}^e_{\mathrm{mat}}=\int_{S_0}\mathbf{B}^{\mathrm{T}}\mathbf{R}^{\mathrm{T}}\mathbf{C}\mathbf{R}\mathbf{B}\,\mathrm{d}S\label{19b}\\
&\mathbf{K}^e_{\mathrm{geom}}=\int_{S_0}\left[2\mathbf{B}^{\mathrm{T}}\dfrac{\partial\mathbf{R}^{\mathrm{T}}}{\partial
\mathbf{d}}\mathbf{T}+\mathbf{d}^{\mathrm{T}}\mathbf{B}^{\mathrm{T}}\dfrac{\partial\mathbf{R}^{\mathrm{T}}}{\partial
\mathbf{d}}\mathbf{C} \dfrac{\partial\mathbf{R}}{\partial
\mathbf{d}}\mathbf{B}\mathbf{d}\right.\nonumber\\&\left.+\left(\mathbf{B}^{\mathrm{T}}\mathbf{R}^{\mathrm{T}}\mathbf{C}
\dfrac{\partial\mathbf{R}}{\partial
\mathbf{d}}\mathbf{B}\mathbf{d}+\mathbf{d}^{\mathrm{T}}\mathbf{B}^{\mathrm{T}}\dfrac{\partial\mathbf{R}^{\mathrm{T}}}{\partial
\mathbf{d}}\mathbf{C}\mathbf{R}\mathbf{B}\right)\right]\,\mathrm{d}S\label{19c}
\end{align}
\end{subequations}

In case of small displacements, Eq.\eqref{eq19} reduces to the
standard form of the material contribution to the element stiffness
matrix, Eq.\eqref{19b}. In case of large displacements, the complete
tangent stiffness matrix is composed of four terms, see
Eq.\eqref{19c}. Only the first one, involving the computation of the
cohesive traction vector $\mathbf{T}$ is not symmetric. Therefore, a
nonsymmetric solver has to be used. However, in case of a symmetric
constitutive matrix $\mathbf{C}$ for the interface, as it happens in
case of the same CZM parameters for Mode I and Mode II, we explored
the possibility to neglect the non symmetric contribution to
$\mathbf{K}_g$. The examples discussed in Section 5 will show that
this will not affect the accuracy of the solution and slightly
penalize the convergence rate. The omission of this term, on the
other hand, makes it possible the use of symmetric solvers. This
fact represents an obvious advantage in case of massive
computations.

\section{Material models}

With reference to the continuum, to assess the effect of large
displacements on debonding of thin structures, both a small
deformation and a large deformation versions of a standard
homogeneous isotropic hyperelastic material model are considered in
the sequel. The derivation is here omitted for the sake of brevity.
The readers can refer to \cite{FEAP} for more details.

Regarding the cohesive traction vector $\mathbf{T}$, with the aim of
quantifying the role of the geometric nonlinear effects along the
decohesion process, two different types of interface constitutive
laws are examined.

First, a tension cut-off CZM is considered, with uncoupled Mode I
and Mode II deformation. This type of CZM has the advantage of
allowing for closed form solutions for specific testing
configurations, like for the double cantilever beam test \cite{WH}.
The stiffness of the CZM can be related to the Young's modulus $E$
and to the thickness of the adhesive $h$, i.e.
$k=\sigma_{\max}/l_{\text{nc}}\sim E/h$ where $\sigma_{\max}$ and
$l_{\text{nc}}$ denote the critical traction for damage initiation
and the critical relative displacement, respectively. In this
approach, when the crack sliding or opening displacements overcome a
critical value corresponding to the achievement of the adhesive
strength, $\sigma_{\max}$, the interface suddenly debonds. Since
such critical relative displacements for failure are very small
quantities in applications, the process zone size is expected to be
quite small and limited within a region very close to the real crack
tip, where displacements are moderately small.

The cohesive traction vector $\mathbf{T}=(\tau,\sigma)^{\mathrm{T}}$ reads:
\begin{subequations}\label{eq22}
\begin{align}
\tau=&\tau_{\max}\dfrac{g_{\mathrm{loc,t}}}{l_{\text{tc}}}\\
\sigma=&\sigma_{\max}\dfrac{g_{\mathrm{loc,n}}}{l_{\text{nc}}}
\end{align}
\end{subequations}
where $g_{\mathrm{loc,t}}$ and $g_{\mathrm{loc,n}}$ are
the tangential and normal components of the gap vector
$\mathbf{g}_{\mathrm{loc}}$, whereas $l_{\text{tc}}$ and
$l_{\text{nc}}$ are the critical sliding and opening displacements.
The tangent constitutive matrix stemming from the linearization of
the CZM tractions with respect to the gap vector is:
\begin{equation}\label{eq23}
\mathbf{C}=\left[
             \begin{array}{cc}
               \dfrac{\tau_{\max}}{l_{\text{tc}}} & 0 \\
               0 & \dfrac{\sigma_{\max}}{l_{\text{nc}}} \\
             \end{array}
           \right]
\end{equation}
In this case, in line with the previous arguments, the resulting interface element stiffness matrix is always symmetric.

As second formulation, we consider the polynomial CZM by Tvergaard
\cite{TV} as an example of an interface constitutive relation where
the cohesive traction vector $\mathbf{T}=(\tau,\sigma)^{\mathrm{T}}$
is a nonlinear function of the sliding and opening displacements
with a softening branch after reaching the maximum cohesive
tractions. For the same values of the parameters $\tau_{\max}$,
$\sigma_{\max}$ and of the initial stiffness as for the tension
cut-off model, this CZM has a larger fracture energy and therefore a
more widespread process zone is expected. In this model, the
cohesive tractions are given by
\begin{subequations}\label{eq24}
\begin{align}
\tau=&\tau_{\max}\dfrac{g_{\mathrm{loc},t}}{l_{\text{tc}}}P(\lambda)\\
\sigma=&\sigma_{\max}\dfrac{g_{\mathrm{loc},n}}{l_{\text{nc}}}P(\lambda)
\end{align}
\end{subequations}
where
\begin{subequations}\label{eq25}
\begin{align}
P(\lambda)=&\left\{\begin{array}{ll}
             \dfrac{27}{4}\left(1-2\lambda+\lambda^2\right), & \text{for}\, 0\leq\lambda\leq 1 \\
             0, & \text{otherwise}
           \end{array}\right.\\
\lambda=&\sqrt{\left(\dfrac{g_{\mathrm{loc},n}}{l_{\mathrm{nc}}}\right)^2+\left(\dfrac{g_{\mathrm{loc},t}}{l_{\mathrm{tc}}}\right)^2}
\end{align}
\end{subequations}

For this CZM, the tangent constitutive matrix reads:
\begin{equation}\label{eq26}
\begin{aligned}
\mathbf{C}=&\left[
             \begin{array}{c}
              \tau_{\max}\dfrac{P}{l_{\mathrm{tc}}}+\tau_{\max}\dfrac{g_{\mathrm{loc},t}}{l_{\mathrm{tc}}}\dfrac{\partial P}{\partial\lambda}\dfrac{\partial\lambda}{\partial g_{\mathrm{loc},t}} \\
              \tau_{\max}\dfrac{g_{\mathrm{loc},t}}{l_{\mathrm{tc}}}\dfrac{\partial P}{\partial\lambda}\dfrac{\partial\lambda}{\partial g_{\mathrm{loc},n}}
             \end{array}\right.\\
             &\left.\begin{array}{c}
             \sigma_{\max}\dfrac{g_{\mathrm{loc},n}}{l_{\mathrm{nc}}}\dfrac{\partial P}{\partial\lambda}\dfrac{\partial\lambda}{\partial g_{\mathrm{loc},t}}\\
              \sigma_{\max}\dfrac{P}{l_{\mathrm{nc}}}+\sigma_{\max}\dfrac{g_{\mathrm{loc},n}}{l_{\mathrm{nc}}}\dfrac{\partial P}{\partial\lambda}\dfrac{\partial\lambda}{\partial g_{\mathrm{loc},n}}
               \end{array}
           \right]
\end{aligned}
\end{equation}

\section{Matrix operators for finite element implementation}

This section covers the main features concerning the numerical
implementation of the large displacement interface element
formulation proposed in the previous section. According to the
derivation presented in Section 2, we restrict our attention to the
implementation of the element in a 2D version, although Eqs.(18) and
(19) are the same for 3D problems, provided that a middle surface is
introduced in analogy with the middle line for the 2D case.

Let us to consider a 4 node bilinear interface element, see
Fig.\ref{fig3}. The corresponding shape functions to accomplish the
numerical integration are $N_1=\dfrac{1}{2}(1-\xi)$ and
$N_2=\dfrac{1}{2}(1+\xi)$. Each node has two degrees of freedom, so
that the nodal position and displacement vectors are arranged as:
\begin{subequations}\label{eq20}
\begin{align}
\mathbf{X}&=(X_1,Y_1,X_2,Y_2,X_3,Y_3,X_4,Y_4)^{\mathrm{T}}\\
\mathbf{d}&=(u_1,v_1,u_2,v_2,u_3,v_3,u_4,v_4)^{\mathrm{T}}
\end{align}
\end{subequations}
where $X_i$, $Y_i$ identifies the cartesian coordinates
corresponding to the node $i$ and $u_i$ and $v_i$ stands for the
corresponding displacements along the $X$ and $Y$ directions.

The gap and the traction vectors that characterizes the CZM are
evaluated in correspondence of each integration point, so that:
\begin{subequations}\label{eq20b}
\begin{align}
\mathbf{g}_{\mathrm{loc}}&=(g_{\mathrm{loc,t}},g_{\mathrm{loc,n}})^{\mathrm{T}}\\
\mathbf{T}&=(\tau,\sigma)^{\mathrm{T}}
\end{align}
\end{subequations}

Next, the matrix operators defined in Section 2 to determine the
coordinates and the gaps of the points belonging to the interface
middle line take the form:
\begin{subequations}\label{eq21}
\begin{align}
\mathbf{N}&=\left[
              \begin{array}{cc}
                N_1\mathbf{I}\; & N_2\mathbf{I}
              \end{array}
            \right]\\
\mathbf{M}&=\dfrac{1}{2}\left[
              \begin{array}{cccc}
                \mathbf{I} & \mathbf{0} & \mathbf{0} & \mathbf{I}\\
                \mathbf{0} & \mathbf{I} & \mathbf{I} & \mathbf{0}\\
              \end{array}
            \right]\quad
\mathbf{L}=\left[
              \begin{array}{cccc}
                -\mathbf{I} &  \mathbf{0} & \mathbf{0} & \mathbf{I}\\
                 \mathbf{0} & -\mathbf{I} & \mathbf{I} & \mathbf{0}\\
              \end{array}
            \right]
\end{align}
\end{subequations}
where $\mathbf{0}$ is a $2\times 2$ null matrix and $\mathbf{I}$ is
a $2\times 2$ identity matrix. As previously stated, the CMZ is
evaluated at the local reference system that is defined by the
tangential vector $\mathbf{t}$ and the normal vector $\mathbf{n}$ to
this middle line. Thus, the rotation operator yields:
\begin{equation}
\mathbf{R}=\left[
              \begin{array}{cc}
                t_x & t_y \\
                n_x & n_y \\
              \end{array}
            \right]
\end{equation}
where:
\begin{equation*}
\begin{aligned}
t_x&=n_y=\dfrac{X_2+u_2+X_3+u_3-X_1-u_1-X_4-u_4}{2\|\mathbf{t}\|}\\
t_y&=-n_x=\dfrac{Y_2+v_2+Y_3+v_3-Y_1-v_1-Y_4-v_4}{2\|\mathbf{t}\|}
\end{aligned}
\end{equation*}
and the symbol $\|.\|$ denotes the Euclidean norm of the
corresponding vector. Note that, differing from previous interface
formulations \cite{PW11b}, this operator is evaluated at each
integration point at the element level.

The operator stemming from the third order tensor has to be computed
with care. For the present particular case it renders:
\begin{equation}
\dfrac{\partial\mathbf{R}}{\partial\mathbf{d}}\mathbf{B}\mathbf{d}=\left[
                                                                      \begin{array}{cccccccc}
                                                                      -a & -b & +a & +b & +a & +b & -a & -b \\
                                                                      -b & +a & +b & -a & +b & -a & -b & +a \\
                                                                      \end{array}
                                                                    \right]
\end{equation}
where $a=\dfrac{-N_1u_1-N_2u_2+N_2u_3+N_1u_4}{2\|\mathbf{t}\|}$ and
$b=\dfrac{-N_1v_1-N_2v_2+N_2v_3+N_1v_4}{2\|\mathbf{t}\|}$.

The remaining operations to accomplish are straightforward, and therefore are omitted here for the sake of conciseness.

The algorithmic treatment of the proposed formulation, implemented
by the present authors both in the Finite Element Analysis Program
FEAP and in the commercial software Abaqus as user defined
subroutines, is summarized in the following sequence of main
operations, see Algorithm 1. Note that the external loop over $j$
refers to the numerical integration, whereas the variable $nlgeom$
indicates a flag defined in the input file to select between: (i)
small displacement formulation ($nlgeom = 0$) or (ii) large
displacement formulation ($nlgeom = 1$).

\begin{algorithm}
\KwData{Given: $\mathbf{X}^{e}$, $\mathbf{d}$, $\Delta \mathbf{d}$,
$\Delta t$}
\eIf{nlgeom  $>0$}
{ Update the geometry: $\mathbf{x}^{e}= \mathbf{X}^{e} + \mathbf{d}$\; }
{ $\mathbf{x}^{e} = \mathbf{X}^{e}$\; }
Construct $\mathbf{L}$\;
Loop over the integration points\;
\For{$j=1$ \KwTo $2$ }
{ Evaluate shape functions and derivatives\;
Compute the local basis vectors $[\mathbf{t}, \mathbf{n}]$ and rotation matrix $\mathbf{R}$\;
Construct $\mathbf{N}$\;
Perform  $\mathbf{B} = \mathbf{N}\mathbf{L}$\;
Compute  $\mathbf{R}\mathbf{B}$ $\rightarrow$ $(\mathbf{R}\mathbf{B})^{T}$\;
Evaluate $\mathbf{T}$, $\mathbf{C}$ according to the selected CZM\;}
\eIf{nlgeom  $>0$}{Perform: $\frac{\partial \mathbf{R}}{\partial \mathbf{d}}\mathbf{B}\mathbf{d}$\;
Construct the geometrical stiffness matrix $\mathbf{K}_{\mathrm{geom}}^{e}$, Eq.\eqref{19c}\;
Compute the geometric part of the residual vector $\mathbf{f}_{\mathrm{intf}}^e$, second term of Eq.(15)\;}
{Compute the material part of the residual vector $\mathbf{f}_{\mathrm{intf}}^e$,first term of Eq.(15)\;}
{Evaluate the material contribution to the stiffness matrix $\mathbf{K}_{\mathrm{mat}}^e$, Eq.\eqref{19b}\;}
\eIf{nlgeom $>0$}{$\mathbf{K}^e = \mathbf{K}_{\mathrm{mat}}^e + \mathbf{K}_{\mathrm{geom}}^e$ }
{ $\mathbf{K}^e =\mathbf{K}_{\mathrm{mat}}^e$ }
{Update stiffness matrix and r.h.s. vector\; }
\caption{Numerical implementation of the large displacement interface element.} \label{Algorithm1}
\end{algorithm}

\section{Numerical examples}

In this section, the numerical performance of the proposed element
is illustrated. To this aim, two applications are selected. First,
we investigate the element capabilities through a series of
benchmark problems to highlight the principal capabilities that the
present element formulation incorporates. Second, a structural
application consisting of a peeling test is addressed. Small or
finite displacement formulations for the continuum and for the
interface element are examined, along with two different CZM
formulations.

\subsection{Benchmark tests}

\begin{figure}
\centering \subfigure[Sketch of the test
problem]{\includegraphics[width=.3\textwidth,angle=0]{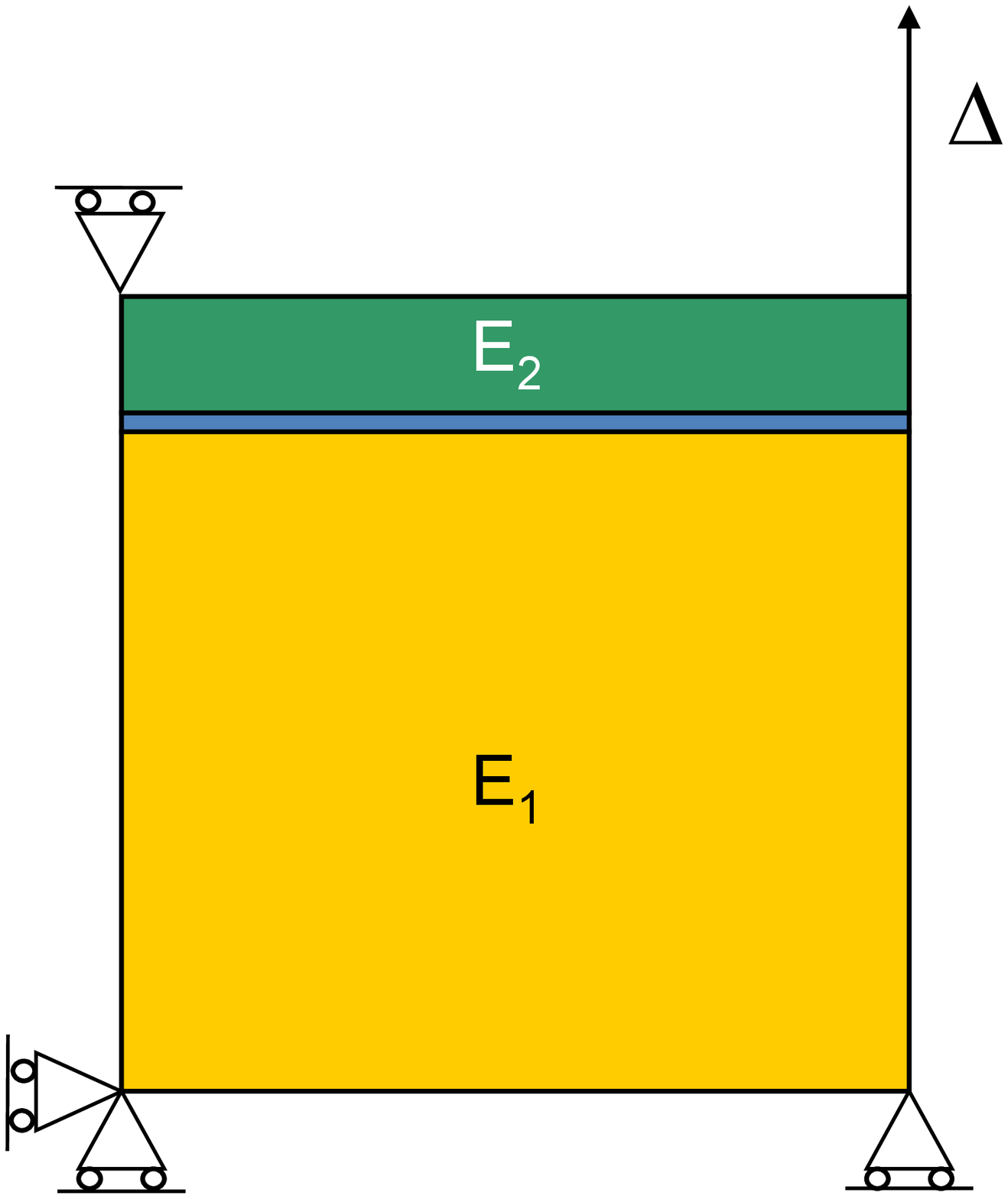}\label{fig4a}}\quad
\subfigure[$\sigma/\sigma{\max}$ vs.
$g_{\mathrm{loc,n}}$]{\includegraphics[width=.4\textwidth,angle=0]{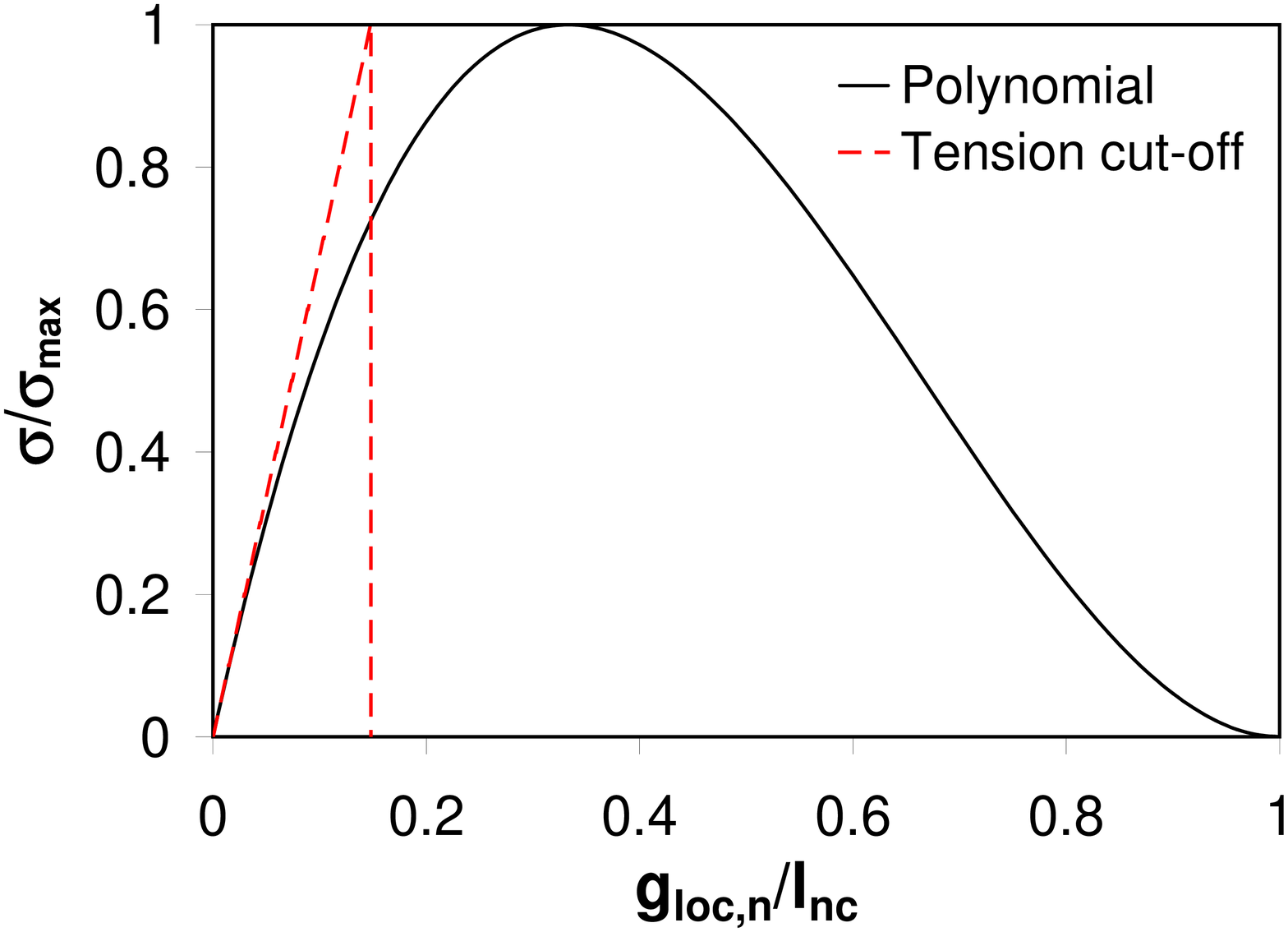}\label{fig4b}}
\caption{Sketch of the geometry of the 2-blocks Mixed Mode test
problem and illustration of the normal cohesive traction-normal gap
CZM relations used in the simulations.}\label{fig4}
\end{figure}

A preliminary test problem shown in Fig.\ref{fig4a} is analyzed in
order to assess the performance of the new interface element for
large displacements as compared to a standard formulation for small
displacement. This benchmark problem aims at investigating the interplay
between geometric and material nonlinearities, an issue not yet
rigorously quantified in the related literature. Therefore, as was
previously stated, we consider small or large deformation
hyperelastic material models for the continuum in order to
investigate the role played by the different interface element
formulations in these two cases. Hence, each numerical test will be
identified by labels $X-Y (Z)$. Whereas $X$ refers to the interface
element formulation and it can be $S$ or $L$ depending on the small
or the large displacement formulation used, the second label $Y$
stands for the constitutive model of the continuum and again it can
be $S$ or $L$ depending on the small or large deformation theory
adopted for the hyperelastic material. The symbol $Z$ in parenthesis
denotes the solver used ($s$ for symmetric solver or $u$ for a non
symmetric solver).

In this benchmark problem, two blocks of lateral size 1 mm and different
heights are discretized by a single finite element each. The lower
block of unit height has a Young's modulus $E_1=500$ MPa in order to
simulate an almost rigid substrate, whereas the upper block has a
height of 0.1 mm and a Young modulus $E_2=5$ MPa to simulate a
highly deformable elastomeric tape or a paper sheet. Both materials
have a vanishing Poisson's ratio and the simulations are conducted
under plane strain assumption. An interface element is placed
between the two blocks and its CZM can have either a tension cut-off
or a polynomial form, see Fig.\ref{fig4b} for the Mode I relations
(Section 3). The values of the CZM parameters $\sigma_{\max}$ and
$\tau_{\max}$ are selected the same for both the tension cut-off and
the polynomial CZMs, to perform a consistent comparison. On the
other hand, the fracture energy of the tension cut-off model is much
smaller than that of the polynomial CZM, as it can be readily
visualized in Fig.\ref{fig4b} from the different area under the
respective traction-separation relations. The value of
$l_{\mathrm{nc}}$ in the tension cut-off model has been selected so
that the tension cut-off curve is tangent to the polynomial CZM at
$g_{\mathrm{loc,n}}=0$.

As far as the boundary conditions are concerned, a non-uniform Mixed
Mode debonding problem is simulated by restraining the first block
at the basis and imposing a linear vertical displacement variation
to the upper side of the second block. This testing configuration
has been chosen because leads to large displacements at the interface
during the deformation process and therefore Mode Mixity \cite{OR}. In other
types of loading, such as uniform Mode I debonding, uniform Mode II
debonding or uniform Mixed Mode debonding, the response of the
interface element would be in fact the same regardless of the small
or large displacement formulation used, since the orientation of the
local frame does not change during the deformation process. These
trends are consistent with the results reported in \cite{BSG07},
although they were based on an approximated large displacement
formulation for the interface element deduced by the kinematics of a
beam element in large displacements and rotations.

\begin{figure}
\centering \subfigure[Tension cut-off
CZM]{\includegraphics[width=.45\textwidth,angle=0]{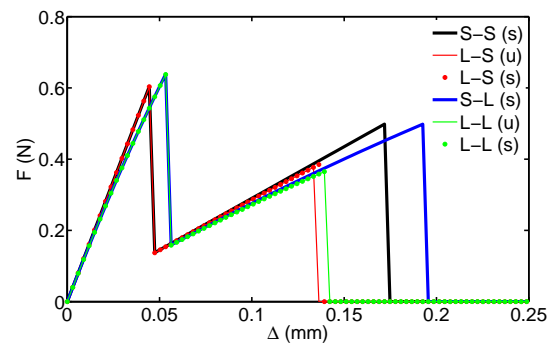}\label{fig5a}}\\
\subfigure[Polynomial
CZM]{\includegraphics[width=.45\textwidth,angle=0]{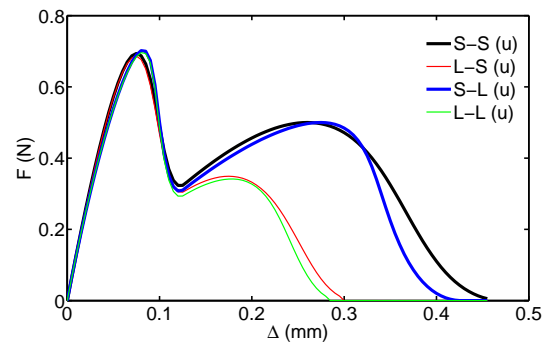}\label{fig5b}}
\caption{Total force vs. imposed displacement for the test problem
in Fig.\ref{fig4a} and different CZM formulations for an interface
tougher in Mode I than in Mode II ($\sigma_{\max}/\tau_{\max}=10$).
$(S-S)$: small displacement interface element \& small deformation
continuum; $(L-S)$: large displacement interface element \& small
deformation continuum; $(S-L)$: small displacement interface element
\& large deformation continuum; $(L-L)$: large displacement
interface element \& large deformation continuum.}\label{fig5}
\end{figure}

The vertical reaction force $F$ in the constrained node is plotted
vs. the imposed displacement $\Delta$ in Fig.\ref{fig5} for an
interface tougher in Mode I than in Mode II, i.e., with
$\sigma_{\max}/\tau_{\max}=10$ and
$l_{\mathrm{nc}}=l_{\mathrm{tc}}=0.1$ mm. The results for the
tension cut-off CZM are shown in Fig.\ref{fig5a} and those for the
polynomial CZM are depicted in Fig.\ref{fig5b}. Examining the curves
related to the tension cut-off CZM, Fig.\ref{fig5a}, we note that,
in case of the small deformation theory for the continuum, curves
labeled $(S-S)$ and $(L-S)$ are almost coincident before the peak
load, i.e., before complete debonding of the first Gauss point of
the interface. Note that the labels $(s)$ and $(u)$ make reference
to the symmetric or unsymmetric character of the formulation. In
this sense, the small-displacement case always leads to a symmetric
stiffness matrix, whereas the large displacement case provides a
general unsymmetric formulation, in which the role of the term that
break the symmetry (see Eq.(20c)) is examined.

After the peak load, an abrupt reduction in the interface
load-carrying capacity takes place and the structural response is
characterized by a lower stiffness until complete decohesion takes
place. In this respect, the large displacement formulation for the
interface predicts a much lower displacement jump for complete
debonding. A similar difference can be noticed in case of a large
deformation formulation for the continuum, see curves labeled
$(S-L)$ and $(L-L)$. Since this CZM does not consider coupling
between Mode I and Mode II, i.e., the matrix $\mathbf{C}$ is
diagonal, it makes sense to compare the results for the large
displacements interface element formulation by considering the
complete expression of the tangent stiffness matrix and using a non
symmetric solver or the approximate symmetric expression and using a
symmetric solver. As it can be seen from Fig.\ref{fig5a}, the
results are coincident.

In case of the polynomial CZM, the results have the same trend as
for the tension cut-off, just with much smoother curves during the
debonding process. In light of the previous arguments, it is worth
noting that for a given assumption regarding the kinematics of the
continuum, the use of a large displacement formulation for the
interface instead of its small displacement counterpart has a
predominant effect on the softening branch of the $F-\Delta$
response. In this case, since the matrix $\mathbf{C}$ is not
symmetric due to the expression of the CZM, a non symmetric solver
has been always used and the full expression for the geometric
stiffness matrix has been retained in the computations.

\begin{figure}
\centering
\subfigure[Tension cut-off CZM]{\includegraphics[width=.45\textwidth,angle=0]{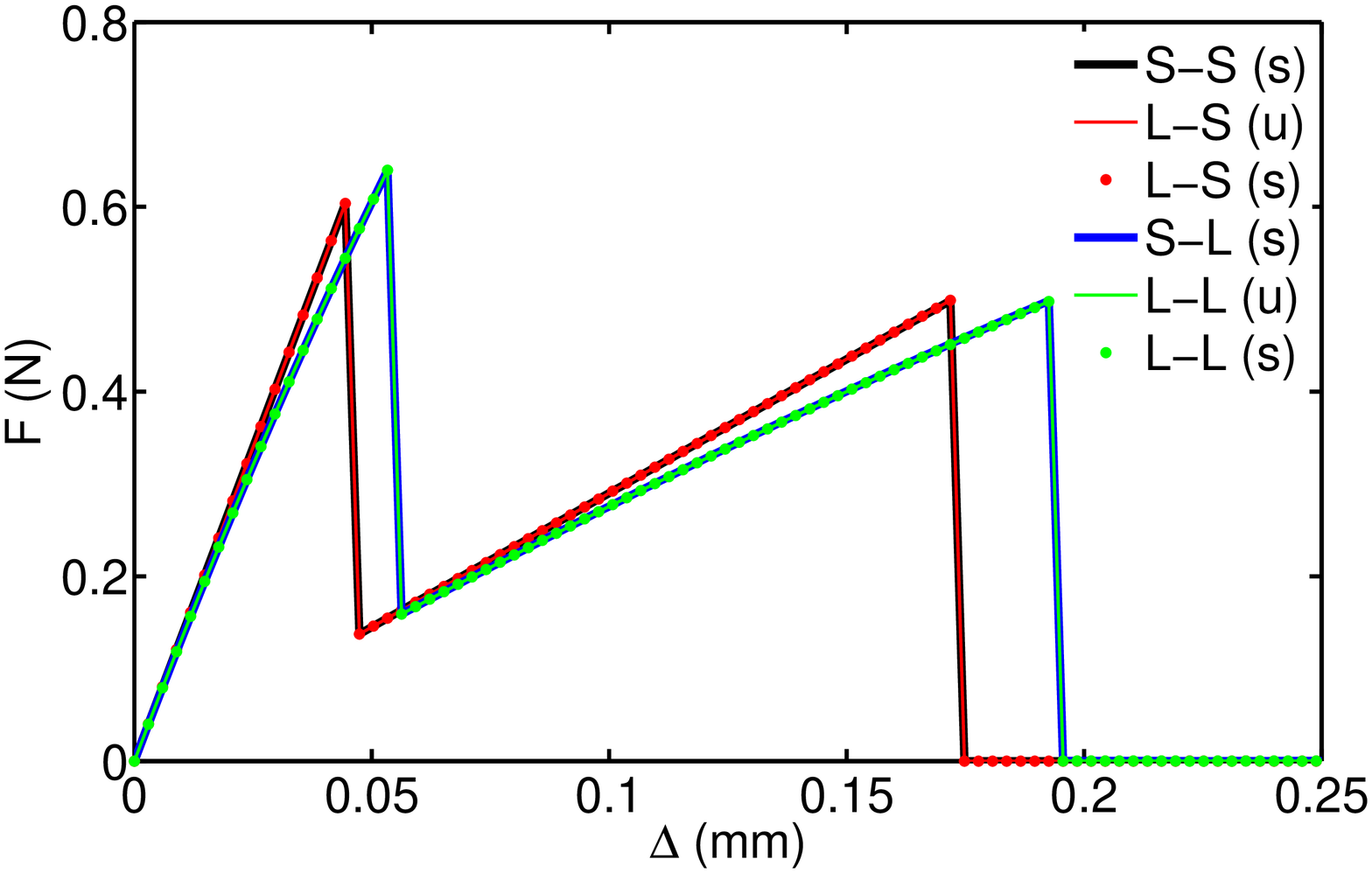}\label{fig6a}}\\
\subfigure[Polynomial CZM]{\includegraphics[width=.45\textwidth,angle=0]{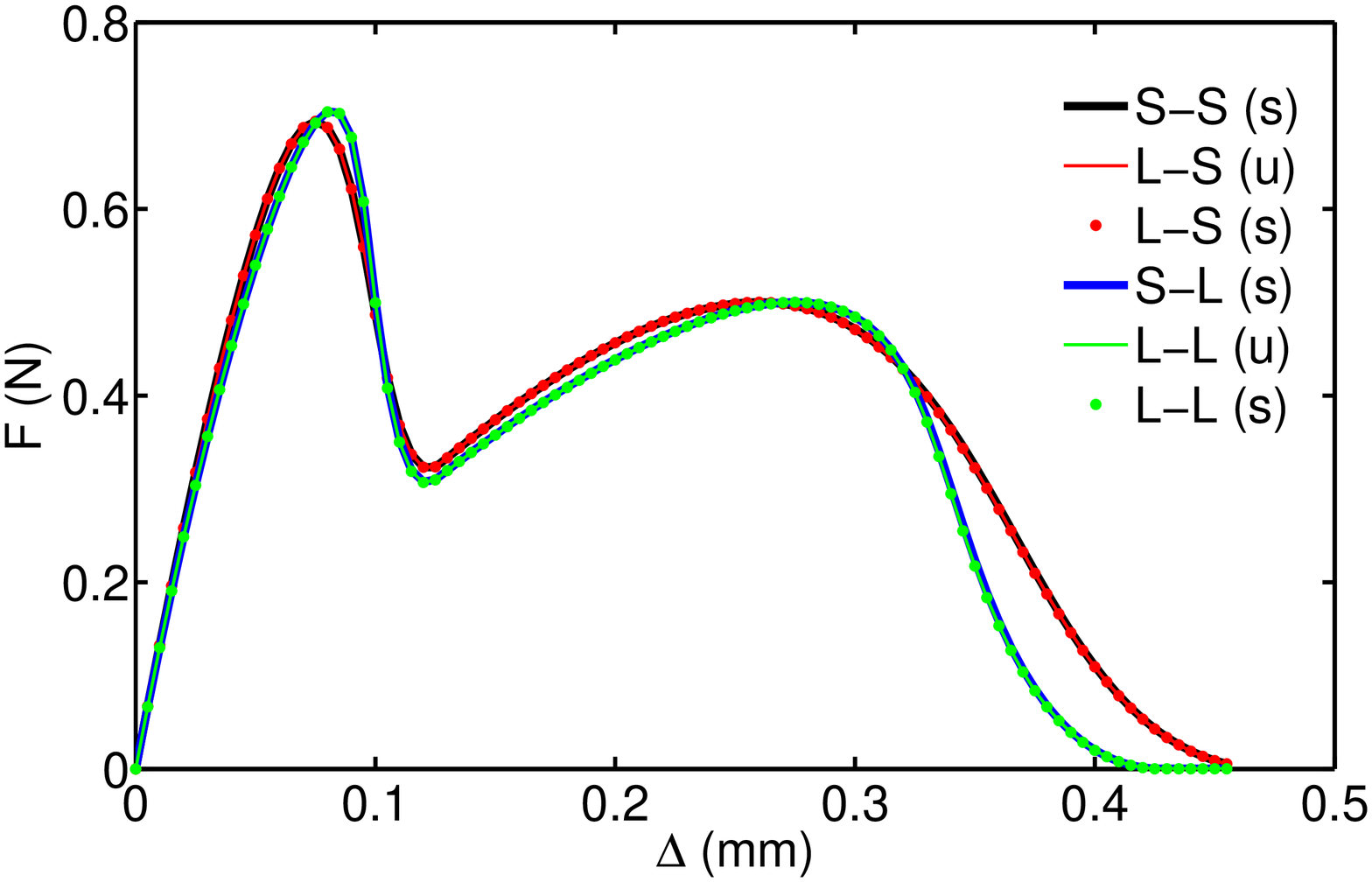}\label{fig6b}}
\caption{Total force vs. imposed displacement for the test problem in Fig.\ref{fig4a} and different CZM formulations for an interface with the same toughness in Mode I and in Mode II ($\sigma_{\max}/\tau_{\max}=1$). $(S-S)$: small displacement interface element \& small deformation continuum; $(L-S)$: large displacement interface element \& small deformation continuum; $(S-L)$: small displacement interface element \& large deformation continuum; $(L-L)$: large displacement interface element \& large deformation continuum.}\label{fig6}
\end{figure}

Examining the element performance in case of an interface with the
same fracture parameters in Mode I and in Mode II
($\sigma_{\max}/\tau_{\max}=1$), see Fig.\ref{fig6}, we find that
the discrepancy between the predictions in case of large or small
interface element formulations are minimal. On the other hand, small
or large displacement formulations for the continuum significantly
affects the post-peak branch. Since in this case the matrix
$\mathbf{C}$ is symmetric for both the CZM formulations, the use of
the complete non symmetric tangent stiffness matrix or its symmetric
version by neglecting the non symmetric contribution to its
geometric component have been compared and the results are again
coincident.

\begin{figure}
\centering
\subfigure[Tension cut-off CZM]{\includegraphics[width=.45\textwidth,angle=0]{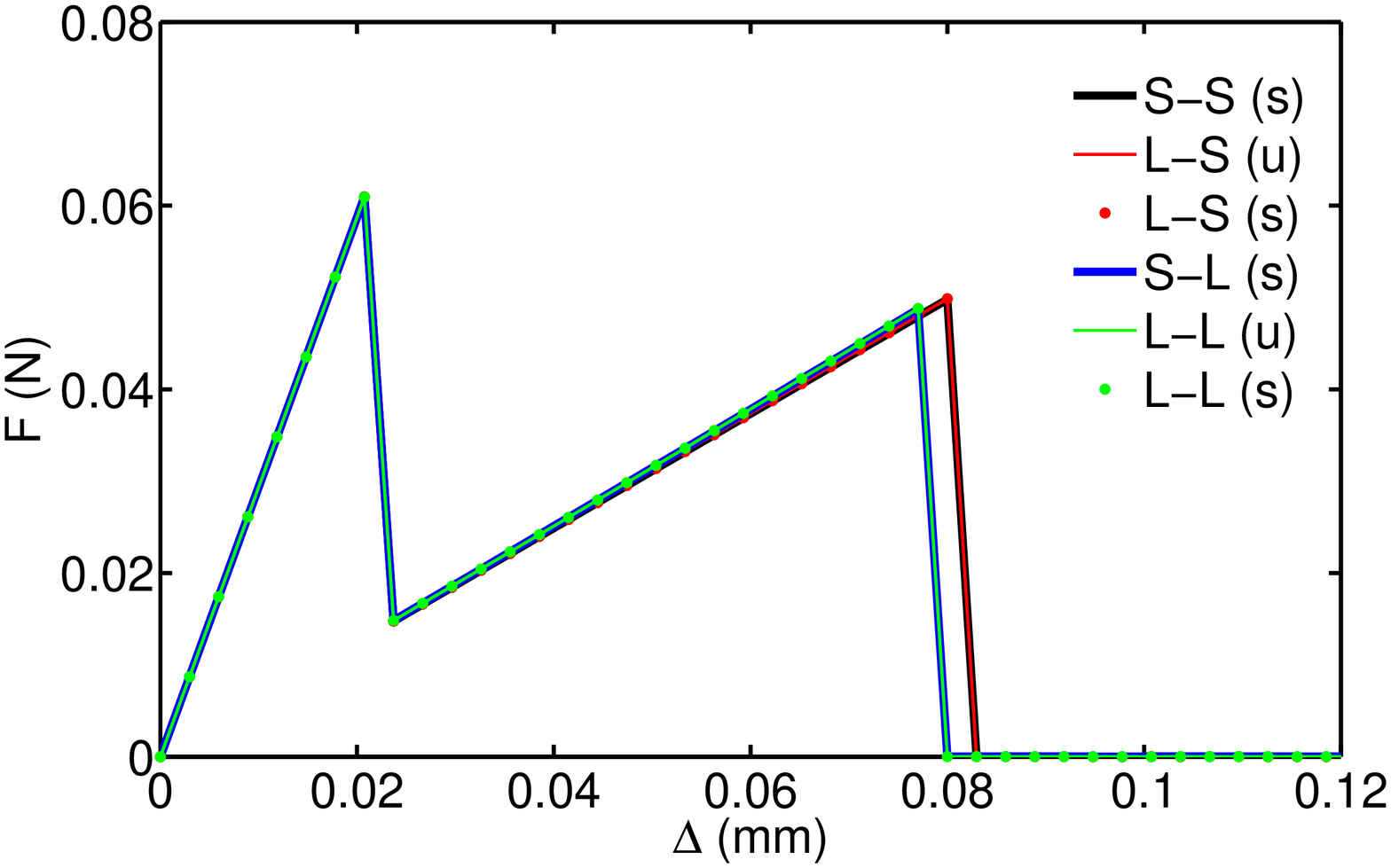}\label{fig7a}}\\
\subfigure[Polynomial CZM]{\includegraphics[width=.45\textwidth,angle=0]{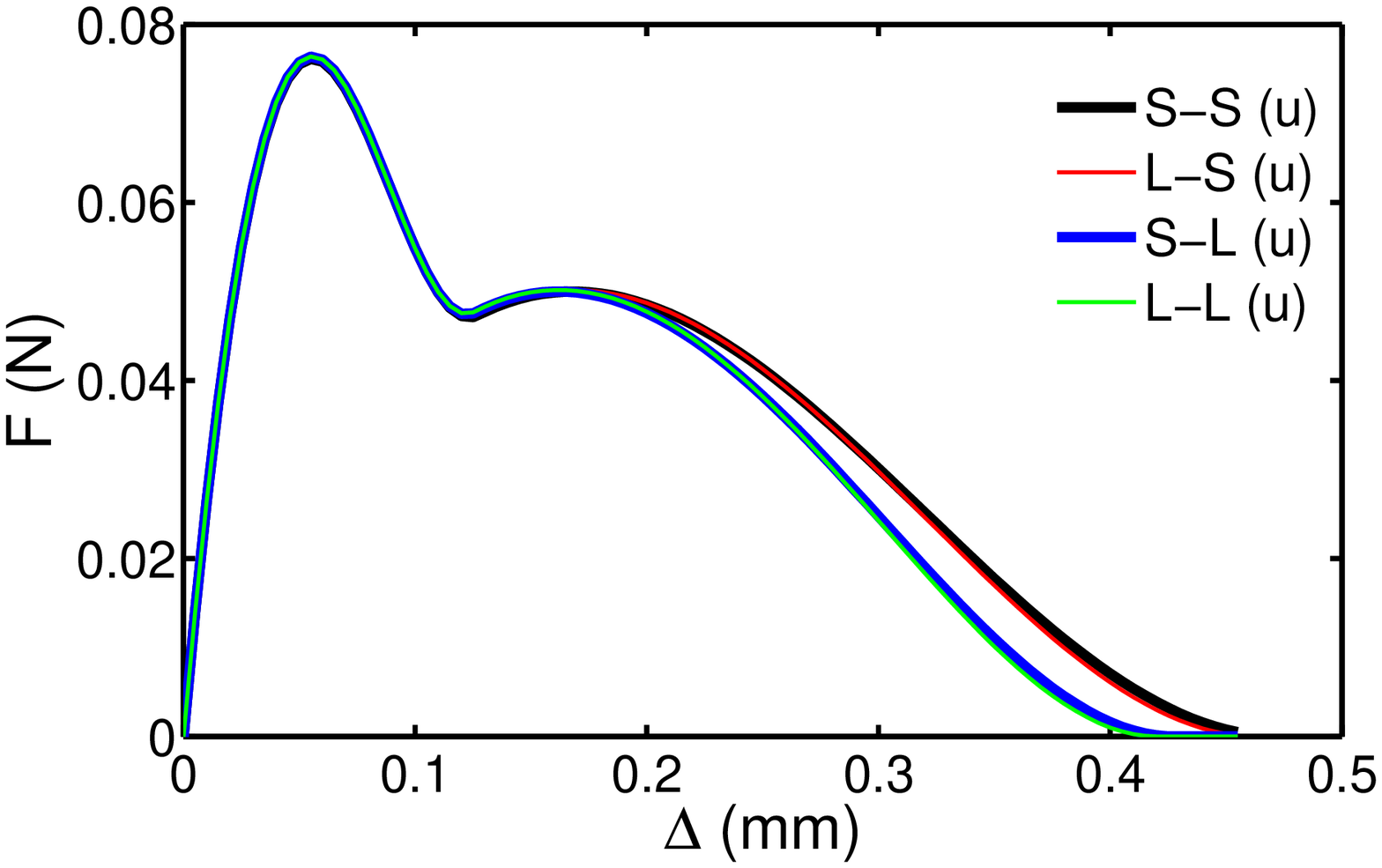}\label{fig7b}}
\caption{Total force vs. imposed displacement for the test problem in Fig.\ref{fig4a} and different CZM formulations for an interface tougher in Mode II than in Mode I ($\sigma_{\max}/\tau_{\max}=0.1$). $(S-S)$: small displacement interface element \& small deformation continuum; $(L-S)$: large displacement interface element \& small deformation continuum; $(S-L)$: small displacement interface element \& large deformation continuum; $(L-L)$: large displacement interface element \& large deformation continuum.}\label{fig7}
\end{figure}

Finally, the last scenario to be inspected is represented by the
case of an interface much tougher in Mode II than in Mode I
($\sigma_{\max}/\tau_{\max}=0.1$), see Fig.\ref{fig7}. As far as the
choice of the symmetric or the non symmetric solver is concerned,
the same comments to the case when Mode I prevails over Mode II
apply. In this instance, since the loading test is predominantly in
Mode I, we observe a much lower peeling force $F$ in this benchmark
test. Additionally, slight discrepancies between the numerical
predictions using large or small interface element formulations are
noticed.

Therefore, it is possible to draw the practical conclusion that the
large displacement formulation for the interface should be primarily
used in case of applications with $\sigma_{\max}>\tau_{\max}$, as,
e.g., in fibrilation problems where the shear strength of cellulose
or polymeric fibrils is almost negligible as compared to their axial
strength.

\subsection{Structural application: peeling test and comparison with experiments}

\begin{figure}
\centering
\includegraphics[width=.45\textwidth,angle=0]{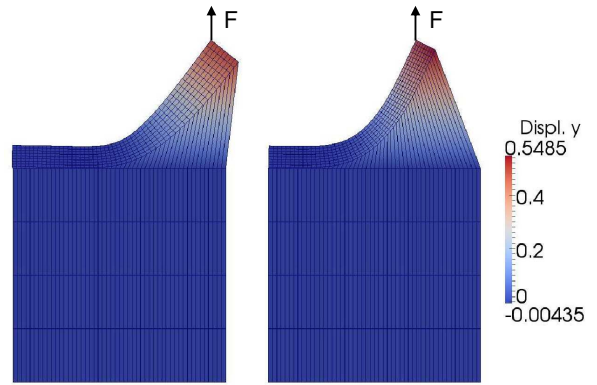}
\caption{Deformed meshes of the peeling test in case of small displacement formulations for the continuum and the interface element (left, $(S-S)$) or large displacement formulations (right, $(L-L)$).}\label{fig8}
\end{figure}

Examining now a structural problem where the large displacement
formulation for the interface element is deemed to be crucial, a
peeling test where a thin layer is pulled from al almost rigid
substrate by the action of a vertical displacement imposed to the
top right corner is considered (see the final deformed shapes in
case of $(S-S)$ or $(L-L)$ formulations in Fig.\ref{fig8}). The
material parameters for the bulks and for the CZMs (tension cut-off
and polynomial CZMs) are the same as in the previous example,
considering the case of an interface tougher in Mode I than in Mode
II ($\sigma_{\max}/\tau_{\max}=10$, where the large displacement
formulation for the interface element was found to significantly
differ from the small displacement one. A non symmetric solver and
the complete expression for the tangent stiffness matrix are used.

\begin{figure}
\centering
\subfigure[Tension cut-off CZM]{\includegraphics[width=.45\textwidth,angle=0]{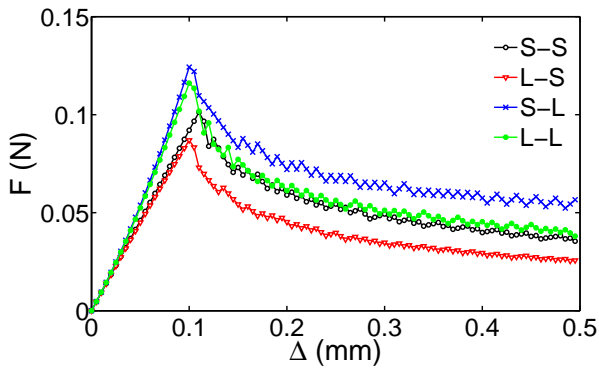}\label{fig9a}}\quad
\subfigure[Polynomial CZM]{\includegraphics[width=.45\textwidth,angle=0]{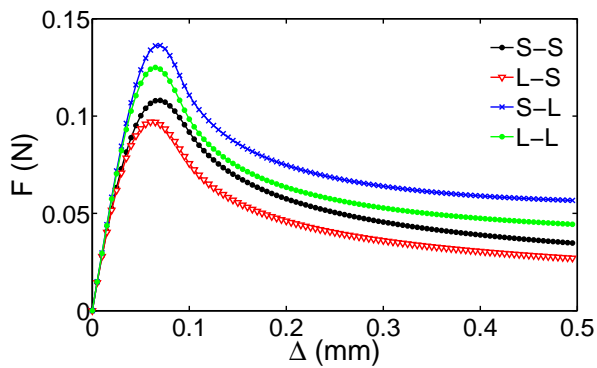}\label{fig9b}}
\caption{Total force vs. imposed displacement for the peeling test in Fig.\ref{fig7} and different CZM formulations for an interface tougher in Mode I than in Mode II ($\sigma_{\max}/\tau_{\max}=10$)). $(S-S)$: small displacement interface element \& small deformation continuum; $(L-S)$: large displacement interface element \& small deformation continuum; $(S-L)$: small displacement interface element \& large deformation continuum; $(L-L)$: large displacement interface element \& large deformation continuum.}\label{fig9}
\end{figure}

The force-displacement curves for different kinematics formulations
are compared in Fig.\ref{fig9}. As a general trend, the large
displacement formulation for the interface element leads to lower
peak loads as compared to its small displacement counterpart, for a
given kinematical model of the continuum. Large differences among
the predictions of the formulations can also be observed as far as
the softening branches are concerned.

Additionally, some comments on the convergence of the formulation
herein proposed have to be added. First, from the numerical point of
view, the interface element formulation was found to be quite
stable, with the appearance of small oscillations in the softening
branches for the peeling test only in case of the tension cut-off
CZM. These effects are caused by the sharp discontinuity in the
traction-gap constitutive relation leading to small jumps in the
load when debonding takes place in a given Gauss point, see
Figs.\ref{fig9a}. These small oscillations disappear in case of the
polynomial CZM, since a softening is included in the interface
constitutive relation.

\begin{figure}
\centering
\includegraphics[width=.45\textwidth,angle=0]{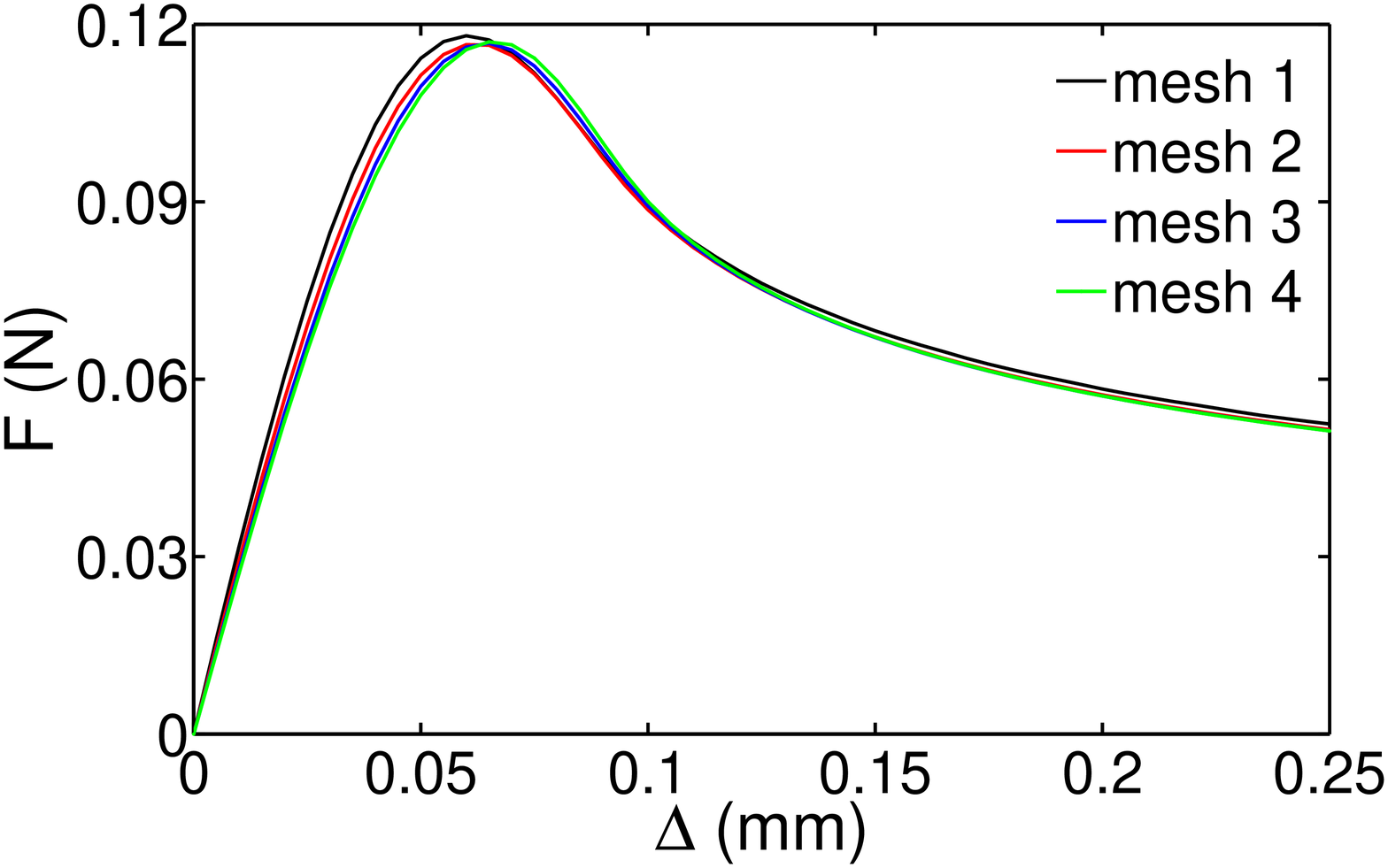}
\caption{Mesh convergence study for the peeling test in
Fig.\ref{fig8}. Mesh 1 corresponds to 25 elements along the
interface, mesh 2 to 50 elements, mesh 3 to 100 elements and mesh 4
to 200 elements.}\label{fig10}
\end{figure}

Second, for the sake of completeness, the mesh convergence of the
method is tested by considering the polynomial CZM and performing
peeling tests as in Fig.\ref{fig9b} for the (L-L) case, with
different mesh refinement for the bulk and the interface in the
horizontal direction. In the coarsest discretization (mesh 1), only
25 elements along the interface are used. Finer meshes with 50
elements (mesh 2), 100 elements (mesh 3) and 200 elements (mesh 4)
along the interface are also considered. The corresponding
load-displacement curves are shown in Fig.\ref{fig10}, where an
excellent mesh-independency even for the coarsest mesh can be
observed.

The predictive capabilities of the proposed formulation are finally
checked against experimental results. To this aim, a $90^\circ$
peeling of a backsheet (0.1 mm thick, Young's modulus 2.8 GPa,
vanishing Poisson's ratio, hyperelastic material) from a glass
substrate (4 mm thick, Young's modulus 73 GPa, vanishing Poisson's
ratio, linear elastic material) is simulated by modelling the
adhesive response of the Epoxy Vynil Acetate (EVA) interlayer via
the polynomial CZM used in the previous examples. The parameters to
be identified are the peak cohesive traction $\sigma_{\max}$ and the
fracture energy $G_{\mathrm{Ic}}$, which is proportional to the
critical opening displacement $l_{\mathrm{nc}}$. The same parameters
for Mode I and Mode II deformation are used and the symmetric
formulation of the interface element for large displacement analyses
is adopted. The test is conducted under plane strain conditions.
Mesh and boundary conditions are analogous to those displayed in
Fig.8, although the upper layer is much thinner in the present
problem. Four finite elements are used to discretize the backsheet
through its thickness and 200 finite elements are used along the
interface, considering an initial bonded length of 50 mm.

For this test, essential to ascertain the reliability of backsheet
bonding in photovoltaic systems, experimental results are reported
in \cite{solmat}. Since the material parameters and the exact
dimensions corresponding to the experimental force-peel extension
curve are not listed in \cite{solmat}, we use values conforming to
the standard materials used in PV production, see \cite{JSA}.
Another source of uncertainty regards the way the peeling extension
is measured, since no details are provided in \cite{solmat}. In the
numerical simulation we predict the peeling extension via the
location of the fictitious crack tip position. Keeping in mind that,
alternatively, the position of the real crack tip could be used,
this choice makes a certain difference especially in the pre-peak
branch of the force-peel extension curve.

\begin{figure}
\centering
\includegraphics[width=.45\textwidth,angle=0]{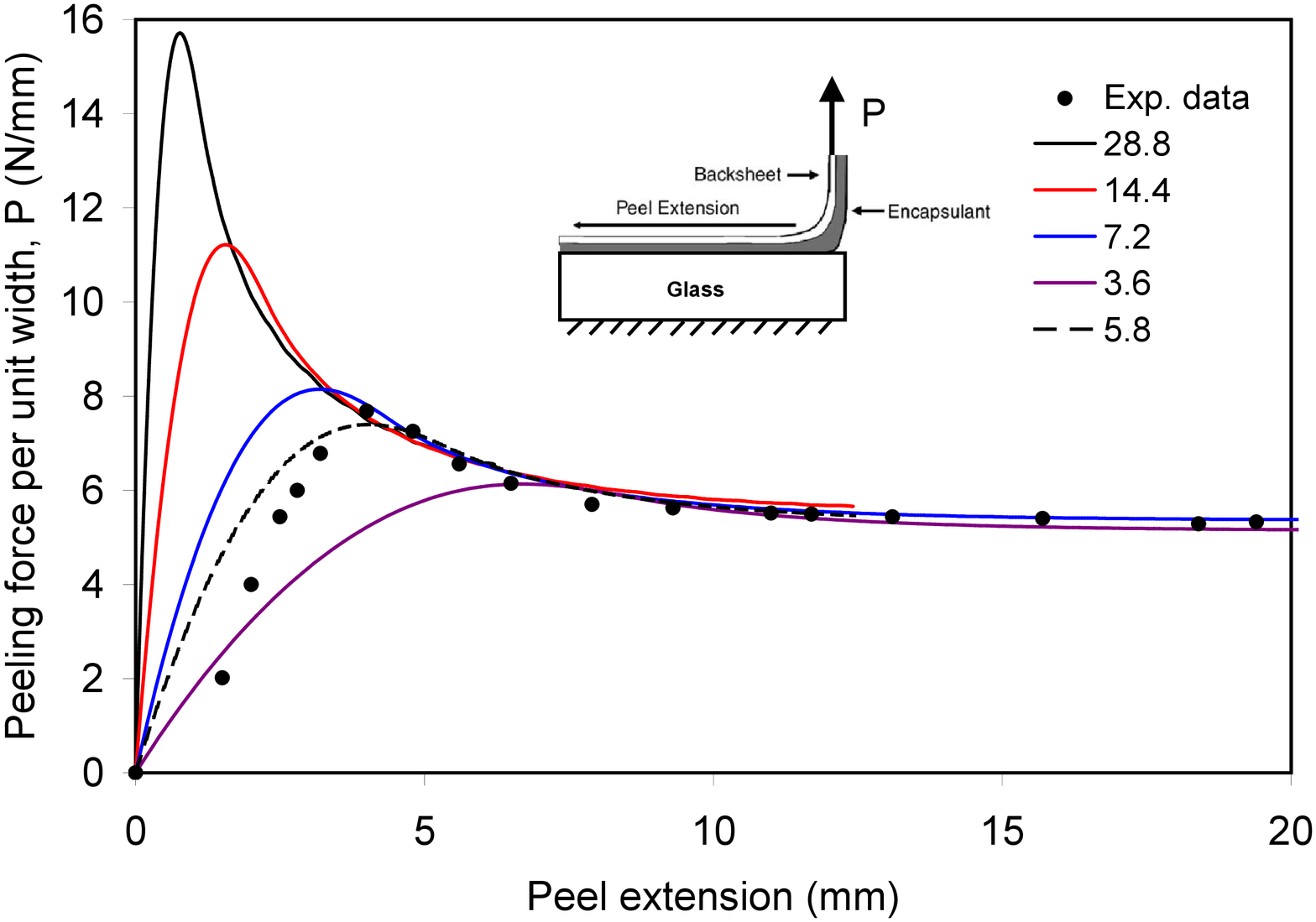}
\caption{Peeling of a backsheet from a glass substrate: numerical
vs. experimental results taken from \cite{solmat}.}\label{fig11}
\end{figure}

Numerical results are shown in Fig.\ref{fig11} for
$G_{\mathrm{Ic}}=5.4$ N/mm, which corresponds to the steady-state
peeling force measured in experiments. A series of curves obtained
by varying $\sigma_{\max}$ in the range from 3.6 to 28.8 N/mm are
displayed and are used to identify the value of $\sigma_{\max}$
which provides the best agreement with experiments. In the present
case, we found $\sigma_{\max}\sim 5.8$ N/mm, corresponding to the
black dashed curve in Fig.\ref{fig11}. The numerical methods is able
to predict the stead-state peeling force very well, in excellent
agreement with experiments. The pre-peak response, which has the
highest degree of inaccuracy in the real tests, is in any case
reasonably well reproduced.

\section{Conclusions and outlook}

In this paper, a consistent derivation of an interface element for
large displacements applications has been proposed.

The present theory finds its variational basis in the interface
contribution to the Principle of Virtual Work of the whole
mechanical system. Our present work, differing from alternative
formulations presented in the literature, furnishes a consistent
derivation of the interface model involving large deformations.
Particularly, the cohesive model herein developed takes into account
the full finite kinematics in which the material and the geometrical
contributions to the element stiffness matrix are clearly
determined.

The corresponding finite element discretization of the interface
model has been accomplished based on a linear two-dimensional
zero-thickness interface element for which the fundamental operators
and the implementation details have been addressed. The compact and
consistent theoretical derivation allows its straightforward
generalization to different orders of the kinematic interpolation
and to 3D topologies.

Numerical applications using two different cohesive interface laws
and with small or finite deformation kinematic assumptions for the
continuum have been examined in order to assess the interface
element performance. Particularly, concerning the interface
constitutive models, two laws have been selected: (1) the so-called
tension cut-off model, that assimilates a quasi-brittle behavior of
the interface, and (2) the polynomial based Tvergaard model that was
adopted for simulating a ductile interface. The numerical results
have proven the applicability of the interface element proposed
especially in terms of its satisfactorily numerical convergence in
achieving equilibrium solutions, along with a minimal mesh
sensitivity. The predictive character of the method has been
demonstrated through the simulation of a peeling test of a backsheet
from a glass substrate, in which the ability of the formulation to
capture the nonlinear character of the experimental trend is
noteworthy.

In closing, we would like to emphasize that the developed model is
particularly promising in addressing real situations undergoing
large displacements, which commonly take place in a wide range of
engineering and biomechanical applications. This fact has been
evidenced in the peeling simulations included in this investigation
where the role of finite displacements has been highlighted. In case
of problems involving thin layers with an interface much tougher in
Mode I than in Mode II, as in fibrilation problems, the proposed
interface element for large displacements is recommended to be used
instead of its small displacement counterpart, to avoid a
significant overestimation of the peeling force, as shown in the
examples discussed in the present study.

\vspace{1em} \addcontentsline{toc}{section}{Acknowledgements}
\noindent\textbf{Acknowledgements} \vspace{1em}

\noindent The research leading to these results has received funding
from the European Research Council under the European Union's
Seventh Framework Programme (FP/2007-2013) / ERC Grant Agreement n.
306622 (ERC Starting Grant ``Multi-field and multi-scale
Computational Approach to Design and Durability of PhotoVoltaic
Modules" - CA2PVM; PI: Prof. M. Paggi). JR would like to acknowledge
the financial support by the above ERC Starting Grant, supporting
his visiting period at IMT Lucca during March-April 2014, and also
the German Federal Ministry for Education and Research (BMBF) for
supporting the project ``Microcracks: Causes and consequences for
the long-term stability of PV-modules" (2012-2014).

\bibliographystyle{unsrt}

\begin{thebibliography}{99}

\bibitem{giB}
Barenblatt GI (1962) The mathematical theory of equilibrium cracks in brittle fracture. Adv Appl Mech 7:55--129.

\bibitem{EGGP}
Elices M, Guinea G, G\'{o}mez J, Planas J (2002) The cohesive zone
model: advantages, limitations and challenges. Eng Fract Mech
69:137--63.

\bibitem{zamm}
Carpinteri A, Paggi M (2012) Modelling strain localization by
cohesive/overlapping zones in tension/compression: Brittleness size
effects and scaling in material properties. Z Angew Math Mech
92:829--840.

\bibitem{PWCM}
Paggi M, Carpinteri A, Wriggers P (2012) Special issue on
computational methods for interface mechanical problems. Comp Mech
50:269--271.

\bibitem{ngu}
Ngu D, Park K, Paulino G, Huang Y (2010) On the constitutive
relation of materials with microstructure using a potential-based
cohesive model for interface traction. Eng Fract Mech 77:1153--1174.

\bibitem{aH}
Hillerborg A (1990) Fracture mechanics concepts applied to moment
capacity and rotational capacity of reinforced concrete beams. Eng
Fract Mech 35:233--240.

\bibitem{AC}
Carpinteri A (1989) Post-peak and post-bifurcation analysis on cohesive crack propagation. Eng Fract Mech 32:265--278.

\bibitem{AC1}
Carpinteri A (1989) Cusp catastrophe interpretation of fracture
instability. J Mech Phys Solids 37:567--582.

\bibitem{AC2}
Carpinteri A (1989) Softening and snap-back instability in cohesive
solids. Int J Numer Meth Eng 28:1521-1537.

\bibitem{PW11}
Paggi M, Wriggers P (2011) A nonlocal cohesive zone model for finite thickness interfaces – Part I: mathematical formulation and validation with molecular
dynamics. Comput Mat Sci 50:1625--1633.

\bibitem{allix}
Allix O, Corigliano A (1996) Modeling and simulation of crack propagation in mixed-modes interlaminar fracture specimens. Int J Fract, 77:111--140.

\bibitem{camanho}
Turon A, Camanho PP, Costa J, D\'{a}vila CG (2006) A damage model for the simulation of delamination in advanced composites under variable-mode loading. Mech Mater, 38:1072--1089.

\bibitem{reinoso}
Reinoso J, Bl\'{a}zquez A, Estefani A, Par\'{\i}s F, Ca\~{n}as J, Ar\'{e}valo E, Cruz F (2012) Experimental and three-dimensional global-local finite element analysis of a composite
component including degradation process at the interfaces. Composites: Part B, 43:1929--1942.

\bibitem{HS}
Hattiangadi A, Siegmund T (2004) A thermomechanical cohesive zone model for bridged delamination cracks. J Mech Phys Solids 52:533--566.

\bibitem{OBG}
Ozdemir I, Brekelmans WAM, Geers MGD (2010) A thermo-mechanical cohesive zone model. Comput Mech 26:735--745.

\bibitem{SP}
Sapora A, Paggi M (2013) A coupled cohesive zone model for transient analysis of thermoelastic interface debonding. Comput Mech 53:845--857.

\bibitem{BSG08}
van den Bosch MJ, Schreurs PJG, Geers MGD (2008) Identification and characterization of delamination in polymer coated metal sheet. J Mech Phys Solids 56:3259--3276.

\bibitem{Petal}
Paggi M, Lehmann E, Weber C, Carpinteri A, Wriggers P, Schaper M (2013) A numerical investigation of the interplay between cohesive cracking and plasticity in polycrystalline materials. Comput Mat Sci 77:81--92.

\bibitem{PW12}
Paggi M, Wriggers P (2012) Stiffness and strength of hierarchical polycrystalline materials with imperfect interfaces. J Mech Phys Solids 60:557--572.

\bibitem{CPZ}
Carpinteri A, Paggi M, Zavarise G (2008) The effect of contact on
the decohesion of laminated beams with multiple microcracks. Int J
Solids Struct 45:129--143.

\bibitem{sacco1}
Alfano G, Sacco E (2006) Combining interface damage and friction in
a cohesive-zone model. Int J Num Meth Eng, 68:542--582.

\bibitem{borino}
Parrinello F, Failla B, Borino G (2009) Cohesive–frictional
interface constitutive model. Int J Solids Struct, 46:2680--2692.

\bibitem{gao}
Yao H, Gao H (2007) Multi-scale cohesive laws in hierarchical materials. Int J Sol Struct 44:8177--8193.

\bibitem{SB}
Schellekens J, de Borst R (1993) On the numerical integration of
interface elements. Int J Numer Meth Eng, 36:44--66.

\bibitem{pande}
Pande GN, Sharma KG (1979) On joint/interface elements and
associated problems of numerical ill-conditioning. Int J Numer Anal
Meth Geomech, 3:293-300.

\bibitem{alfano}
Alfano G, Crisfield MA (2001) Finite element interface models for
the delamination analysis of laminated composites: mechanical and
computational issues. Int J Numer Meth Eng, 50:1701--1736.

\bibitem{borst}
de Borst R (2003) Numerical aspects of cohesive-zone models. Eng
Fract Mech, 70:1743--1757.

\bibitem{remmers}
Remmers JJC, de Borst R, Needleman A (2003) A cohesive segments
method for the simulation of crack growth. Comp Mech, 31:69--77.

\bibitem{OP99}
Ortiz M, Pandolfi A (1999) Finite deformation irreversible cohesive elements for three-dimensional crack-propagation analysis. Int J
Num Meth Engng 44:1267--1282.

\bibitem{RRD02}
Roychowdhury S, Arun Roy Y, Dodds RH (2002) Ductile tearing in thin aluminium panels: experiments and analyses using large-displacement
3-D surface cohesive elements. Eng Frac Mech 69:983--1002.

\bibitem{QCA01}
Qiu Y, Crisfield MA, Alfano G (2001) An interface element formulation for the simulation of delamination with buckling. Eng Frac Mech 68:1755--1776.

\bibitem{BSG07}
van den Bosch MJ, Schreurs PJG, Geers MGD (2007) A cohesive zone model with a large displacement formulation accounting for interfacial fibrilation. European J Mech A/Solids 26:1--19.

\bibitem{FBK13}
Fleischhauer R, Behnke R, Kaliske M (2013) A thermomechanical interface element formulation for finite deformations. Comput Mech 52:1039--1058.

\bibitem{BSG08b}
van den Bosch MJ, Schreurs PJG, Geers MGD (2008) On the development of a 3D cohesive zone element in the presence of large deformations. Comput Mech 42:171--180.

\bibitem{BW}
Bonet J, Wood R (1997) Nonlinear Continuum Mechanics for Finite Element Analysis. Cambridge University Press.

\bibitem{FEAP} Zienkiewicz OC, Taylor RL (2000) The Finite Element Method. Butterworth-Heinemann, Woburn, MA, 5th Edition, Vol. I.

\bibitem{WH} Williams J, Hadavinia H (2002) Analytical solutions for cohesive zone models. J Mech Phys Solids 50:809--825.

\bibitem{TV} Tvergaard V (1990) Effect of fiber debonding in a whisker-reinforced metal. Mat Sci Engng A 107:23--40.

\bibitem{PW11b}
Paggi M, Wriggers P (2011) A nonlocal cohesive zone model for finite thickness interfaces - Part II: FE implementation and application to polycrystalline materials. Comp Mat Sci 50:1634--1643.

\bibitem{OR}
Rabinovitch O (2008) Debonding analysis of fiber-reinforced-polymer
strengthened beams: Cohesive zone modeling versus a linear elastic
fracture mechanics approach. Eng Fracture Mech 75:2842--2859

\bibitem{solmat}
Jorgensen GJ, Terwilliger KM, DelCueto JA, Glick SH, Kempe MD,
Pankow JW, Pern FJ, McMahon TJ (2006) Moisture transport, adhesion,
and corrosion protection of PV module packaging materials. Solar
Energy Materials \& Solar Cells 90:2739--2775.

\bibitem{JSA}
Paggi M, Kajari-Schr\"{o}der S, Eitner U (2011) Thermomechanical
deformations in photovoltaic laminates. The Journal of Strain
Analysis for Engineering Design, 46:772--782.


\end{thebibliography}

\end{document}